\documentclass[reprint, amsmath,amssymb,aps,nofootinbib]{revtex4-1}
\usepackage{graphicx}
\usepackage{dcolumn}
\usepackage{bm}

\newcommand{\be}{\begin{equation}}
\newcommand{\ee}{\end{equation}}
\newcommand{\ba}{\begin{eqnarray}}
\newcommand{\ea}{\end{eqnarray}}
\newcommand{\bef}{\begin{figure}}
\newcommand{\eef}{\end{figure}}
\newcommand{\nn}{\nonumber}

\newcommand{\al}{\alpha}

\newcommand{\si}{\sigma}

\newcommand{\la}{\lambda}
\newcommand{\Lam}{\Lambda}

\newcommand{\ra}{\rightarrow}

\newcommand{\lef}{\Lam_\text{eff}}

\begin{document}

\preprint{APS/123-QED}

\title{The fate of a universe driven by a linear potential}

\author{Ricardo Z. Ferreira}
\email{rferreira@icc.ub.edu}
\affiliation{%
Departament de F\'isica Qu\`antica i Astrofis\'ica \& Institut de Ci\`encies del Cosmos (ICCUB), Universitat de Barcelona, Mart\'i i Franqu\`es 1, 08028 Barcelona, Spain
}%

\author{Pedro P. Avelino}
\email{pedro.avelino@astro.up.pt}
\affiliation{Instituto de Astrof\'isica e Ci\^encias do Espa\c co, Universidade do Porto, CAUP, Rua das Estrelas, PT4150-762 Porto, Portugal \\
Centro de Astrof\'isica da Universidade do Porto, Rua das Estrelas, , PT4150-762 Porto, Portugal and \\
Departamento de F\'isica e Astronomia, Faculdade de Ci\^encias, Universidade do Porto, Rua do Campo Alegre 687, PT4169-007 Porto, Portugal}

\begin{abstract}
We study the proposal to solve the coincidence problem in the non-local version of the vacuum energy sequestering mechanism by means of a scalar field in a linear potential. We show that there is no solution in the theory compatible with observations if one requires the scalar field to drive the present period of acceleration and the collapse.

\end{abstract}

\maketitle

%\tableofcontents

\section{\label{sec:level1} Introduction}

In this note we study the recent proposal of addressing the coincidence problem in the context of the vacuum energy sequestering mechanism (VES) \cite{Kaloper:2013zca, Kaloper:2014dqa, Avelino:2014nqa}. This mechanism proposes a solution to the cosmological constant ($\Lambda$) problem by a modification of General Relativity (GR) which includes non-local terms\footnote{See \cite{Kaloper:2015jra} for a local version of the theory.} such that the Einstein tensor is shielded from vacuum contributions to all loops (excluding graviton loops). In the context of this mechanism a new solution to the coincidence problem was proposed in \cite{Kaloper:2014fca} by means of a scalar field with a linear potential. In this letter we show, however, that  there is no cosmologically viable solution in the theory consistent with the current observational constraints on the curvature parameter $k$ and on the dark energy equation of state parameter $w_\text{DE}$ \cite{Ade:2015xua}.
The paper is organized as follows. In section II we review the dynamics of a linear potential in a FRW Universe. In section III we discuss the VES and show, both analytically and numerically, that a linear potential driving the present stage of acceleration and the collapse in the context of the VES does not have any observationally viable solution.

\section{Linear Potential and Dark Energy}

Consider the Lagrangian density given by
\begin{eqnarray}
{\cal L}= \sqrt{-g}\left[\frac{M_p^2}{2} R + \partial_\mu \phi \partial^\mu \phi + V(\phi) \right]
\end{eqnarray} where $\phi$ is a scalar field with a linear potential $V(\phi) = m^3 \phi$ and $M_p$ is the reduced Planck mass. For a homogeneous scalar field the equation of motion is 
\be \label{5}
\ddot{\phi} + 3H \dot \phi +m^3=0 \, ,
\ee
where $H=\dot a/a$ is the Hubble rate, $a$ is the scale factor and the dot denotes a derivative with respect to the cosmic time (t).
When the Universe is dominated by a perfect fluid with constant equation of state parameter $w$ the scale factor evolves in time as $a=a_0 (t/t_0)^p$, where $p=2/(3(1+w))>1/3$, and $\phi(t)$ evolves as
\be \label{6}
\phi(t)=\phi_i+ \frac{C }{(1-3p) t^{3p-1}} - \frac{m^3 t^2}{2(3p+1)} \,,
\ee
where $\phi_i$ and $C$ are two integration constants. The second term is diluted with the expansion and cannot play an important role today. Otherwise, it would have been even more important in the past and would have spoiled the matter, $w=0$, and radiation, $w=1/3$, domination regimes. Thus, we set $C=0$.

The observational constraints on the dark energy equation of state parameter \cite{Ade:2015xua} $w_\text{DE}+1=(\dot{\phi}^2/2-V(\phi))/(\dot{\phi}^2/2+V(\phi))+1 \lesssim 0.1$ require $\dot \phi^2/2 \ll V(\phi)$ which implies that $m^3/\phi_i \ll H^2$. Therefore, in order to explain the present period of acceleration $\phi_i$ should satisfy\footnote{To be more precise one should also take into account the kinetic term of $\phi$ but that gives only small corrections to $\phi_i$.}
\be \label{8.2}
\phi_i m^3 \simeq \Omega_\text{DE} \rho_0\, ,
\ee
where $\rho_0=3 H_0^2 M_p^2$ is the total energy density and $\Omega_\text{DE}\simeq 0.69 $ \cite{Ade:2015xua} is the dark energy density parameter both evaluated today. 

In the dark energy stage $\phi$ slowly rolls down the potential until $H^2 \simeq m^{3}/M_p$. At that point it gains speed, $\phi$ becomes negative and, in roughly an Hubble time, $\rho_\text{tot}= 3H^2 M_p^2 \rightarrow 0$ and a turnover occurs. Due to present observational constraints, the influence of other components like matter, radiation or curvature is, at this point, very much diluted and cannot interfere with the dynamics.
When $H=0$ the acceleration is negative so the Universe starts contracting dominated by the kinetic energy of $\phi$.

\section{Vacuum energy sequestering}

In this section we briefly review the proposal to solve the $\Lambda$ problem  suggested in \cite{Kaloper:2013zca}. The proposal consists in a modification of GR described by the action
\ba \label{1}
{\cal S}&=& \int d^4 x \sqrt{-g} \left[ \frac{M_p^2}{2} R-\Lambda + \lambda^4 {\cal L} \left(\la^{-2} g^{\mu \nu}, \Phi \right) \right]  \nn \\ &&+  \,   \si \left(\frac{\Lambda}{\la^4 \mu^4}\right), 
\ea
where $R$ is the Ricci scalar, ${\cal L}$ is the Lagrangian of matter and $\Lambda$ contains all the contributions to $\Lambda$ and acts, jointly with $\la$, as a Lagrange multiplier. The associated Einstein equations, after fixing the global constraints, are given by
\be \label{2}
M_p^2 G_{\mu \nu} = T_{\mu \nu} - \frac{g_{\mu \nu}}{4}  \left< T^\al_\al \right>, 
\ee
where $T_{\mu \nu}$ is the stress-energy tensor and $\left< T^\al_\al \right>$ is a {\it historical average} of the trace of $T_{\mu \nu}$ over the history of space-time
\be \label{3}
\left< T^\al_\al \right> \equiv \frac{ \int d^4 x \sqrt{-g} \, T^\al_\al}{ \int d^4x \sqrt{-g}}.
\ee
The key feature of the action defined in Eq. (\ref{1}) is the fact that the gravitational equations are protected from $\Lambda$ to all loop levels (excluding graviton loops). However, there is still a residual contribution $\left< T^\al_\al \right>$ which acts as an effective $\Lam$ ($\lef$). This contribution depends on the full history of the Universe, which has to be large and old in order to sufficiently  dilute $\lef$. In particular, the Universe has to be finite (in space and time) in order for such averages to be well defined. This last property requires
the present stage of acceleration to be transient, the Universe to collapse and the presence of a positive spatial curvature.

Although this proposal addresses the $\Lambda$ problem it does not solve, directly, the coincidence problem. In \cite{Kaloper:2014fca} the authors argued that a scalar field ($\phi$) in a linear potential is a technically natural way to cause the collapse of the Universe, needed by the theory, and that the VES mechanism {\it chooses} initial conditions such that the present dark energy stage starts "now". The argument is based on the exact shift symmetry in the presence of a linear potential. Namely, the impact in $T_{\mu \nu}$ of any shift in $\phi$, $\phi \ra \phi + \phi_0$, is compensated by the corresponding shift in $\left< T^\al_\al \right>/4$ such that Eq. (\ref{2}) remains unchanged. Due to this symmetry, it was shown that if there is a solution it is possible to shift $\phi$ by the right amount such that $\left< T^\al_\al \right>=0$. It was then argued by the authors that such solutions require initial conditions for $\phi$ which would solve the coincidence problem.

We will show that such solutions are not cosmologically viable, i.e., that they do not correspond to universes compatible with current observations. First we show that for $m$ and $k$ consistent with current bounds on $w_\text{DE}+1<0.1$ and $|\Omega_k|=(3kM_p^2/a_0^2)/\rho_0<5 \times 10^{-3}$ \cite{Ade:2015xua}, respectively, the gauge condition $\lef=0$ cannot be satisfied and also give numerical evidence for that. Then, we construct explicit solutions of the theory and find that, as anticipated, they correspond to dynamics where the curvature becomes dominant before the scalar field and so are not compatible with our current knowledge of the universe.
\subsection{Historical Average}
Solutions in the context of the VES require
\be
\left< R \right>= \left< H^2 + \frac{\ddot{a}}{a} +\frac{k}{a^2} \right > =0.
\ee
On the other hand, in order to address the coincidence problem the parameters of the theory should satisfy
\be \label{14}
\rho_\phi |_0 +\lef  = \Omega_\text{DE} \rho_0,
\ee 
where $\rho_0$ is the total energy density, $\rho_\phi|_0$ the energy density of $\phi$ (both evaluated today) and
\be \label{15}
\lef =\frac{\left< T^\al_\al \right>}{4} = \left< \frac{\dot \phi ^2}{4} -  m^3 \phi + (3w-1) \frac{\rho_w}{4} \right>,
\ee
is the historical average of the trace of the stress energy tensor including all matter sources. The term $\rho_w \propto a^{-3(1+w)}$ is the energy density of a generic perfect fluid.
Due to the shift symmetry in the presence of a linear potential, an interachange of $\phi_i$ and $\Lambda$ leaves the theory invariant. Therefore, following the arguments of \cite{Kaloper:2013zca} we work in the gauge where $\lef=0$. Then, $\rho_\phi |_0 = \Omega_\text{DE} \rho_0$ and
\begin{eqnarray} \label{gauge condition}
\left< \frac{\dot \phi ^2}{4} -  m^3 \phi -\frac{\rho_m}{4} \right> =0\, ,
\end{eqnarray}
where we have fixed $w=0$. 

Note that if we start with $\lef=0$ then, at the end of the dynamics, we should ensure that this global constraint is preserved. If it is not we should look for different parameters $m,k$ such that $\lef$ and $\left< R \right> \rightarrow 0$. $\phi_i$ is already fixed by observations so there is no freedom to add a constant to the initial conditions. As we will show, for all parameters $m,k$ compatible with observations, it is not possible to have $\lef=0$.

In what follows we consider a matter field, curvature and $\phi$ although adding any other perfect fluid with $-1/3< w <1$ would lead to the same conclusions.

Let us start by looking at the integrals in the historical average. The denominator is just the space-time volume $\text{Vol}_4= \int d^4x \sqrt{-g} = \text{Vol}_3 \int dt \,a^3 $
where $\text{Vol}_3$ is the spatial volume of the Universe which is finite for $k>0$.
In a Universe which starts in a bang and ends in a crunch $\text{Vol}_4$ converges when $a \ra 0 $ and so it will be dominated by the epoch around the turnover at which $a$ is maximal.

Regarding the numerator in Eq. (\ref{3}), for a perfect fluid
\ba \label{19}
\int d^4x \sqrt{-g} \rho &=& \rho_0 \text{Vol}_3 \int da a^{(1-3w)/2} \nn \\
&\propto& \begin{cases} a^{3/2(1-w)} , \,w \ne 1 \\  \log(a), \, w=1 \end{cases}.
\ea
$\phi$ can be treated as a perfect fluid with $w\simeq-1$ until the turnover and $w\simeq1$ in the collapsing stage, where the kinetic energy of $\phi$ dominates the energy density. 
Matter and curvature are diluted during the dark energy stage by the expansion of the universe, hence, their contribution to the historical integrals are much smaller than the contributions from $\phi$, apart from a short period of time around the turnover. After the turnover, the historical average in eq. (\ref{15}) grows logarithmically because $w_\phi \simeq 1$.

We now look at the evolution of the time integrals appearing in the numerator of the historical averages as a function of time, i.e.
 \begin{eqnarray}
 \lef^{[t]} \equiv \int_{t_i}^t dt' \, a^3 \,T_\alpha^\alpha/4\, ,
 \end{eqnarray}
where $t$ goes from the initial time $t_i$ to the final time $t_f,$ after the turnover, where appropriate physical cutoffs, $\rho_\text{cutoff}\lesssim M_p^4$, are set.  
Just like the denominator in eq. (\ref{3}), during the dark energy stage the numerator grows in time, in absolute value, until the turnover. The term in $\dot{\phi}$ also grows logarithmically in the collapse. Because the integral is a growing function of $t$, exponential during the dark energy dominated stage, then its sign, and, in particular, its zeros, can be estimated by the sign and zeros of the integrand. Therefore, disregarding the logarithmic growth during the collapse, the integrals get most of its contibution at the turnover where $\rho_\text{tot} = 0$, i.e., when
\begin{eqnarray} \label{at the bounce}
m^3 \phi_* = -\rho_{m*} - \frac{\dot{\phi}^2_*}{2} + \frac{3 k M_p^2}{a_*^2}. 
\end{eqnarray}
where the $*$ denotes quantities evaluated at the turnover. 
Therefore, using eq. (\ref{at the bounce}), we can estimate the numerator in $\lef$ to be
\begin{eqnarray}
 &\int_{t_i}^{t_*} dt' \, a^3 \left[\frac{\dot{\phi}^2}{4} - m^3 \phi -\frac{\rho_m}{4} \right] \simeq t_* a_*^3 \left[\frac{\dot{\phi}^2_*}{4} - m^3 \phi_* -\frac{\rho_{m*}}{4}\right] & \nn \\ &=\frac{3}{4} t_* a_*^3 \left[ \dot{\phi}^2_* + \rho_{m*}- \frac{4 k M_p^2}{a_*^2} \right] \gtrsim \frac{3}{4} t_* a_*^3 \dot{\phi_*}^2& ,
\end{eqnarray}
where in the last line we used the fact that, due to the observational constraints on $k$, $4 k M_p^2/a_*^2 \ll \dot{\phi}^2_*$ as we can also see in fig. (\ref{1}) where we show the time evolution of these quantities for the maximal curvature allowed by observations. 
After the turnover $\phi$ remains negative and the kinetic energy of $\phi$ dominates the energy density so the previous integral remains positive and, in particular, grows logarithmically. Therefore, $ \lef^{[t_f]}$ is always positive and larger than the typical energy density at the turnover.
This statement is independent of the precise value of $m$ and $k$ as long as:
\begin{itemize}
\item $m$ and $k$ satisfy the observational constraints;
\item $\phi_i$ is fixed by the amount of dark energy density today.	
\end{itemize}
Therefore, there is no cosmologically viable solution in this context.

\paragraph*{{\bf Numerics:}}
We further substantiate these arguments numerically. In particular, the analytical argument we used to show that $\lef^{[t_f]} \neq 0$ assumes that until the turnover the time integrals can be well approximated by the integrand times the upper limit of integration because they are fast growing functions. In the numerical part we go beyond such approximation reaching the same conclusions.
\bef 
\begin{centering} 
	\includegraphics[scale=0.6]{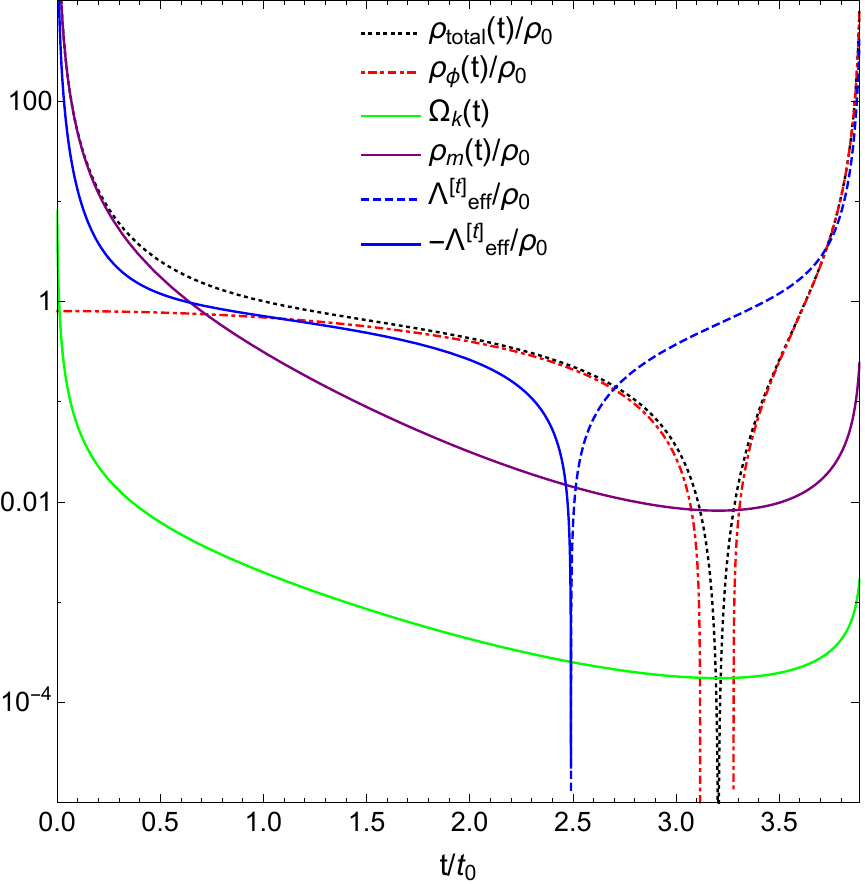}
	\caption{Time evolution of $\rho_\text{tot}$ (green), $\rho_\phi$ (yellow) and $\lef^{[t]}/\rho_0$ (blue, straight when positive and dashed when negative) for $m^3=0.6 \rho_0/M_p$ ($w_\text{DE}=0.1$) and $\Omega_k=2\times10^{-3}$. \label{fig1}} 
	\par\end{centering} 
\eef
As in the previous paragraphs, numerically we consider matter, curvature and the scalar field and evolve the second Friedmann equation, 
\begin{eqnarray}
 M_p^2 \dot{H}= \frac{kM_p^2}{a^2}-\frac{\rho_m}{2}-\frac{\dot{\phi}^2}{2}\, ,
\end{eqnarray}
starting with a zero cosmological constant.
Initial conditions are fixed well inside the matter dominated regime where we impose that $\dot \phi$ satisfies Eq. (\ref{6}). The initial and final time are fixed by a cutoff energy $\rho_\text{cutoff} \ll M_p^4$. When the universe contracts and reaches $\rho_\text{cutoff}$ we evaluate $\lef$ and $\left< R \right>$. We consider cosmologically viable solutions to be those which have $\Lambda_\text{eff} + \rho_\phi |_0 \simeq \Omega_\text{DE} \rho_0$.

In Fig. \ref{fig1} we plot the time evolution of $\rho_\text{tot}$, $\rho_\phi$ and $\lef^{[t]}$ for $m^3=0.6\, \rho_0/M_p$ ($w_\text{DE}\simeq 0.1$) and $\Omega_k=2\times10^{-3}$. As expected, $\lef^{[t]}$ becomes positive before the turnover and stays like that until the collapse.
In Fig. \ref{fig2} we plot instead $\left< R \right>/(3H_0^2)$ and $\lef^{[t_f]}/\rho_0$ as a function of $m$. We chose $\Omega_k$ to be $10^{-3}$ although any other value compatible with observations is almost indistinguishable from the plotted points. Again we verify that $\lef^{[t_f]}/\rho_0 >0$ showing that the gauge $\lef = 0$ does not exist. In particular the value of $\lef^{[t_f]}$ can be of the same order of magnitude of $\rho_0$, depending on $m$. This means that there are no solutions as long as observational constraints are satisfied as we can also see from the behavior of $\left< R \right>\nrightarrow 0$. The two quantities plotted grow, roughly, $\propto m^3$ which is expected as they get most of its support at the turnover where the characteristic energy of the universe is $\rho_* \propto m^3 M_p$ as already noted in \cite{Kaloper:2014dqa}.
\bef 
\begin{centering} 
	\includegraphics[scale=0.6]{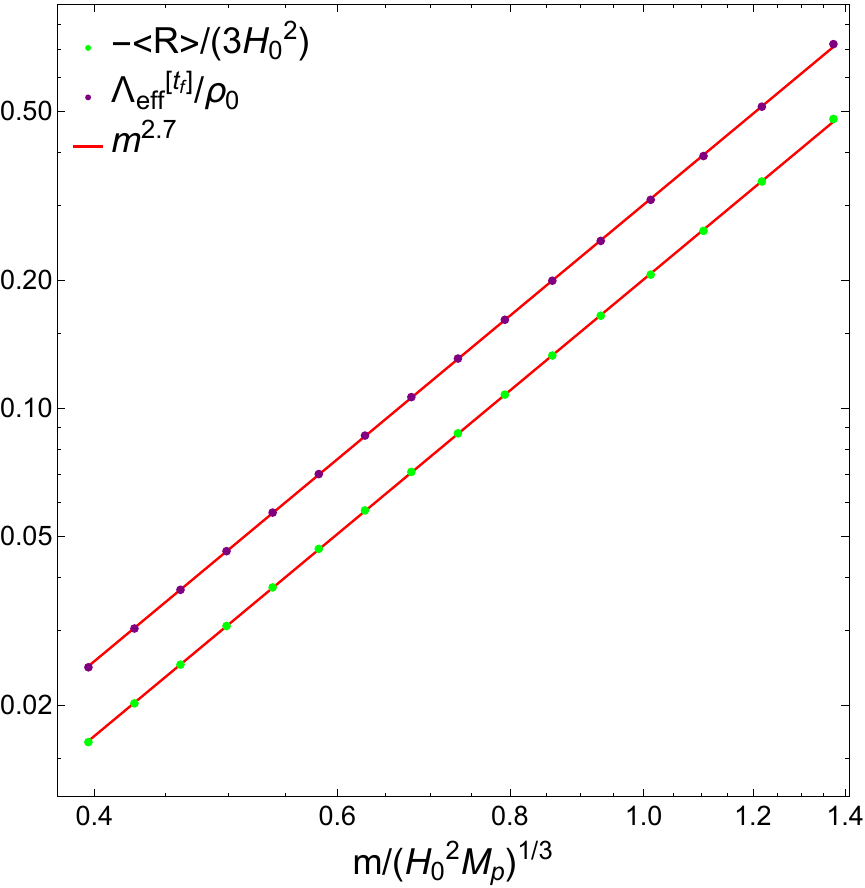}
	\caption{$-\left< R \right>/(3H_0^2)$ and $\lef^{[t_f]}/\rho_0$ evaluated at the cutoff time as a function of $m$ for $\Omega_k=10^{-3}$. This window of masses corresponds to $w_\text{DE} +1 \simeq \left[ 10^{-4}, 0.2  \right] $. The red lines are proportional to $m^{2.7}$. \label{fig2}} 
	\par\end{centering} 
\eef

Finally, we also looked for explicit solutions of the theory without imposing observational bounds. Using the same numerical procedure we increased the value of $k$ until we find a change in the sign of both $\lef$ and $\left< R \right>$. Then, a given $m$ can be chosen such that both quantities converge to zero. The solutions obtained correspond to universes where the curvature density becomes relevant before the scalar field and triggers a collapse. Naturally, such solutions are not cosmologically viable, in agreement with our findings.

\section{Conclusion}

We have studied the proposed solution to the coincidence problem in the context of the vacuum energy sequestering mechanism \cite{Kaloper:2014dqa}. In the non-local version of that proposal it was suggested that a linear potential could explain the present dark energy stage and drive the collapse of the Universe needed in that theory. However, we have provided analytical and numerical evidence that there is no viable solution compatible with observations.
These results do not apply to the local version of the theory \cite{Kaloper:2015jra}.
\linebreak

\paragraph*{{\bf Acknowledgments:}}
RZF would like to thank the Lundbeck foundation, the Danish
National Research Foundation (DNRF90) and Spanish MINECO under MDM-2014-0369 of I$\Lambda$UB (Unidad de Excelencia ''Maria de Maeztu'') for financial support. PPA acknowledges the support of Funda\c c\~ ao para a Ci\^encia e a Tecnologia (FCT) through the research grant UID/FIS/04434/2013. This
paper benefited from the PPA participation on the COST action CA15117 (CANTATA), supported by COST (European Cooperation in Science and Technology)
\bibliographystyle{apsrev4-1}
\bibliography{CoincProblVacSeq}

\begin{thebibliography}{6}%
\makeatletter
\providecommand \@ifxundefined [1]{%
 \@ifx{#1\undefined}
}%
\providecommand \@ifnum [1]{%
 \ifnum #1\expandafter \@firstoftwo
 \else \expandafter \@secondoftwo
 \fi
}%
\providecommand \@ifx [1]{%
 \ifx #1\expandafter \@firstoftwo
 \else \expandafter \@secondoftwo
 \fi
}%
\providecommand \natexlab [1]{#1}%
\providecommand \enquote  [1]{``#1''}%
\providecommand \bibnamefont  [1]{#1}%
\providecommand \bibfnamefont [1]{#1}%
\providecommand \citenamefont [1]{#1}%
\providecommand \href@noop [0]{\@secondoftwo}%
\providecommand \href [0]{\begingroup \@sanitize@url \@href}%
\providecommand \@href[1]{\@@startlink{#1}\@@href}%
\providecommand \@@href[1]{\endgroup#1\@@endlink}%
\providecommand \@sanitize@url [0]{\catcode `\\12\catcode `\$12\catcode
  `\&12\catcode `\#12\catcode `\^12\catcode `\_12\catcode `\%12\relax}%
\providecommand \@@startlink[1]{}%
\providecommand \@@endlink[0]{}%
\providecommand \url  [0]{\begingroup\@sanitize@url \@url }%
\providecommand \@url [1]{\endgroup\@href {#1}{\urlprefix }}%
\providecommand \urlprefix  [0]{URL }%
\providecommand \Eprint [0]{\href }%
\providecommand \doibase [0]{http://dx.doi.org/}%
\providecommand \selectlanguage [0]{\@gobble}%
\providecommand \bibinfo  [0]{\@secondoftwo}%
\providecommand \bibfield  [0]{\@secondoftwo}%
\providecommand \translation [1]{[#1]}%
\providecommand \BibitemOpen [0]{}%
\providecommand \bibitemStop [0]{}%
\providecommand \bibitemNoStop [0]{.\EOS\space}%
\providecommand \EOS [0]{\spacefactor3000\relax}%
\providecommand \BibitemShut  [1]{\csname bibitem#1\endcsname}%
\let\auto@bib@innerbib\@empty
%</preamble>
\bibitem [{\citenamefont {Kaloper}\ and\ \citenamefont
  {Padilla}(2014{\natexlab{a}})}]{Kaloper:2013zca}%
  \BibitemOpen
  \bibfield  {author} {\bibinfo {author} {\bibfnamefont {N.}~\bibnamefont
  {Kaloper}}\ and\ \bibinfo {author} {\bibfnamefont {A.}~\bibnamefont
  {Padilla}},\ }\href {\doibase 10.1103/PhysRevLett.112.091304} {\bibfield
  {journal} {\bibinfo  {journal} {Phys.Rev.Lett.}\ }\textbf {\bibinfo {volume}
  {112}},\ \bibinfo {pages} {091304} (\bibinfo {year} {2014}{\natexlab{a}})},\
  \Eprint {http://arxiv.org/abs/1309.6562} {arXiv:1309.6562 [hep-th]}
  \BibitemShut {NoStop}%
%%CITATION = ARXIV:1309.6562;%%
\bibitem [{\citenamefont {Kaloper}\ and\ \citenamefont
  {Padilla}(2014{\natexlab{b}})}]{Kaloper:2014dqa}%
  \BibitemOpen
  \bibfield  {author} {\bibinfo {author} {\bibfnamefont {N.}~\bibnamefont
  {Kaloper}}\ and\ \bibinfo {author} {\bibfnamefont {A.}~\bibnamefont
  {Padilla}},\ }\href {\doibase 10.1103/PhysRevD.90.084023,
  10.1103/PhysRevD.90.109901} {\bibfield  {journal} {\bibinfo  {journal}
  {Phys.Rev.}\ }\textbf {\bibinfo {volume} {D90}},\ \bibinfo {pages} {084023}
  (\bibinfo {year} {2014}{\natexlab{b}})},\ \Eprint
  {http://arxiv.org/abs/1406.0711} {arXiv:1406.0711 [hep-th]} \BibitemShut
  {NoStop}%
%%CITATION = ARXIV:1406.0711;%%
\bibitem [{\citenamefont {Avelino}(2014)}]{Avelino:2014nqa}%
  \BibitemOpen
  \bibfield  {author} {\bibinfo {author} {\bibfnamefont {P.}~\bibnamefont
  {Avelino}},\ }\href {\doibase 10.1103/PhysRevD.90.103523} {\bibfield
  {journal} {\bibinfo  {journal} {Phys.Rev.}\ }\textbf {\bibinfo {volume}
  {D90}},\ \bibinfo {pages} {103523} (\bibinfo {year} {2014})},\ \Eprint
  {http://arxiv.org/abs/1410.4555} {arXiv:1410.4555 [gr-qc]} \BibitemShut
  {NoStop}%
%%CITATION = ARXIV:1410.4555;%%
\bibitem [{\citenamefont {Kaloper}\ \emph {et~al.}(2015)\citenamefont
  {Kaloper}, \citenamefont {Padilla}, \citenamefont {Stefanyszyn},\ and\
  \citenamefont {Zahariade}}]{Kaloper:2015jra}%
  \BibitemOpen
  \bibfield  {author} {\bibinfo {author} {\bibfnamefont {N.}~\bibnamefont
  {Kaloper}}, \bibinfo {author} {\bibfnamefont {A.}~\bibnamefont {Padilla}},
  \bibinfo {author} {\bibfnamefont {D.}~\bibnamefont {Stefanyszyn}}, \ and\
  \bibinfo {author} {\bibfnamefont {G.}~\bibnamefont {Zahariade}},\ }\href@noop
  {} {\  (\bibinfo {year} {2015})},\ \Eprint {http://arxiv.org/abs/1505.01492}
  {arXiv:1505.01492 [hep-th]} \BibitemShut {NoStop}%
%%CITATION = ARXIV:1505.01492;%%
\bibitem [{\citenamefont {Kaloper}\ and\ \citenamefont
  {Padilla}(2015)}]{Kaloper:2014fca}%
  \BibitemOpen
  \bibfield  {author} {\bibinfo {author} {\bibfnamefont {N.}~\bibnamefont
  {Kaloper}}\ and\ \bibinfo {author} {\bibfnamefont {A.}~\bibnamefont
  {Padilla}},\ }\href {\doibase 10.1103/PhysRevLett.114.101302} {\bibfield
  {journal} {\bibinfo  {journal} {Phys.Rev.Lett.}\ }\textbf {\bibinfo {volume}
  {114}},\ \bibinfo {pages} {101302} (\bibinfo {year} {2015})},\ \Eprint
  {http://arxiv.org/abs/1409.7073} {arXiv:1409.7073 [hep-th]} \BibitemShut
  {NoStop}%
%%CITATION = ARXIV:1409.7073;%%
\bibitem [{\citenamefont {Ade}\ \emph {et~al.}(2015)\citenamefont {Ade} \emph
  {et~al.}}]{Ade:2015xua}%
  \BibitemOpen
  \bibfield  {author} {\bibinfo {author} {\bibfnamefont {P.}~\bibnamefont
  {Ade}} \emph {et~al.} (\bibinfo {collaboration} {Planck}),\ }\href@noop {} {\
   (\bibinfo {year} {2015})},\ \Eprint {http://arxiv.org/abs/1502.01589}
  {arXiv:1502.01589 [astro-ph.CO]} \BibitemShut {NoStop}%
%%CITATION = ARXIV:1502.01589;%%
\end{thebibliography}%

\end{document}